\documentclass[aps,prl,twocolumn,showpacs,superscriptaddress,floatfix]{revtex4-1}
\usepackage{amsmath,amssymb}
\usepackage{graphicx}
\usepackage{epsfig}
\usepackage{color}
\usepackage{rotating}
\usepackage{bm}
\usepackage{setspace}
\begin{document}
\title{Mapping the Electronic Structure of Each Ingredient Oxide Layer of High-$T_\textrm{c}$ Cuprate Superconductor Bi$_2$Sr$_2$CaCu$_2$O$_{8+\delta}$}
\author{Yan-Feng Lv}
\author{Wen-Lin Wang}
\author{Jun-Ping Peng}
\author{Hao Ding}
\author{Yang Wang}
\affiliation{State Key Laboratory of Low-Dimensional Quantum Physics, Department of Physics, Tsinghua University, Beijing 100084, China}
\author{Lili Wang}
\author{Ke He}
\author{Shuai-Hua Ji}
\affiliation{State Key Laboratory of Low-Dimensional Quantum Physics, Department of Physics, Tsinghua University, Beijing 100084, China}
\affiliation{Collaborative Innovation Center of Quantum Matter, Beijing 100084, China}
\author{Ruidan Zhong}
\author{John Schneeloch}
\author{Gen-Da Gu}
\affiliation{Condensed Matter Physics and Materials Science Department, Brookhaven National Laboratory, Upton, New York 11973, USA}
\author{Can-Li Song}
\email[]{clsong07@mail.tsinghua.edu.cn}
\author{Xu-Cun Ma}
\email[]{xucunma@mail.tsinghua.edu.cn}
\author{Qi-Kun Xue}
\email[]{qkxue@mail.tsinghua.edu.cn}
\affiliation{State Key Laboratory of Low-Dimensional Quantum Physics, Department of Physics, Tsinghua University, Beijing 100084, China}
\affiliation{Collaborative Innovation Center of Quantum Matter, Beijing 100084, China}
\date{\today}

\begin{abstract}
Understanding the mechanism of high transition temperature ($T_\textrm{c}$) superconductivity in cuprates has been hindered by the apparent complexity of their multilayered crystal structure. Using a cryogenic scanning tunneling microscopy, we report on layer-by-layer probing of the electronic structures of all ingredient planes (BiO, SrO, CuO$_2$) of Bi$_2$Sr$_2$CaCu$_2$O$_{8+\delta}$ superconductor prepared by argon-ion bombardment and annealing technique. We show that the well-known pseudogap (PG) feature observed by STM is inherently a property of the BiO planes and thus irrelevant directly to Cooper pairing. The SrO planes exhibit an unexpected Van Hove singularity near the Fermi level, while the CuO$_2$ planes are exclusively characterized by a smaller gap inside the PG. The small gap becomes invisible near $T_\textrm{c}$, which we identify as the superconducting gap. The above results constitute severe constraints on any microscopic model for high $T_\textrm{c}$ superconductivity in cuprates.
\end{abstract}
\pacs{74.72.Gh, 68.37.Ef, 74.50.+r, 74.25.Jb}
\maketitle
\begin{spacing}{0.995}
Superconductivity in perovskite-type layered cuprates \cite{bednorz1986possible}, which thus far hold the record for the highest transition temperature ($T_\textrm{c}$), ranks among the most challenging and engaging problems in modern condensed matter physics. Despite nearly three decades' tremendous efforts of research all around the world, the key mechanism behind the Cooper pairing that lies at the heart of high-$T_\textrm{c}$ cuprate superconductors still remains puzzling. Many intriguing phenomena that intertwine with the occurrence of superconductivity have been discovered, leading to a very sophisticated phase diagram of cuprates \cite{keimer2015quantum}. These phenomena include the ubiquitous existence of various sorts of broken-symmetry states (e.g.\ charge density wave, spin density wave and electron nematicity) and the well-known pseudogap (PG) phenomenology, which has been considered a key finding in the research of cuprate superconductivity. A vast amount of experimental and theoretical studies have been devoted to understanding these phenomena themselves and their possible interplay with superconductivity, but so far most of which fell flat.

From the view point of crystal structure, the cuprates consist of superconducting CuO$_2$ layers and charge reservoir building blocks (e.g.\ BiO/SrO in Bi$_2$Sr$_2$CaCu$_2$O$_{8+\delta}$ (Bi-2212)) that stack alternatively along the crystallographic \textit{c}-axis. In Bi-based cuprate superconductors, it is widely thought but empirical that the BiO and SrO block layers are insulating \cite{nieminen2009spectral}. A considerable amount of surface-sensitive measurements, e.g. via angle-resolved photoemission spectroscopy (ARPES) \cite{hashimoto2014energy} and scanning tunneling spectroscopy (STS) \cite{fischer2007scanning}, have been conducted on the vacuum cleaved BiO planes of Bi-based cuprates and contributed largely to the cuprate PG data base. The measurements are generally assumed to reflect the superconducting properties of the CuO$_2$ planes, despite that they are located 4.5 {\AA} beneath the top BiO plane. Yet, such model has not been rigidly tested experimentally thus appears contentious, particularly regarding to the fact that the surficial Bi lattice rather than the buried Cu one was actually visualized in most scanning tunneling microscopy (STM) experiments. STS on the accidentally obtained CuO$_2$ planes shows a vanishing density of states (DOS) around the Fermi level ($E_F$) \cite{kitazawa1993superconducting, murakami1995observation, misra2002atomic}, noticeably differing from the PG on BiO planes \cite{fischer2007scanning}. These facts imply that a systematic experimental study of the electronic properties of each building planes should be carried out to justify the model. The recently discovered high $T_\textrm{c}$ superconductivity in single-unit-cell FeSe/SrTiO$_3$ heterosturctrue \cite{qing2012interface, ge2014superconductivity, bozovic2014new}, in which the superconducting FeSe plane is exposed and directly accessible to STM and ARPES, reveals very simple Fermi surface with nearly isotropic gaps compared to bulk FeSe \cite{liu2012electronic, he2013phase, tan2013interface}. The work strongly suggests that a direct measurement of the electronic spectra of every ingredient planes, particularly the superconducting CuO$_2$ that may have very simple Fermi surface as FeSe/SrTiO$_3$, of cuprates is indispensable for investigating the superconductivity mechanism. This kind of measurement might further allow addressing explicitly the respective role of each ingredient plane, which has been discussed extensively in literatures and is helpful for searching for new superconductors.
\end{spacing}

We report such an \textit{in situ} STM/STS measurement by exposing every planes, namely BiO, SrO and CuO$_2$, of Bi-2212 using state-of-the-art argon-ion bombardment and annealing (IBA) technique. The experiments were carried out in a Unisoku ultrahigh vacuum (UHV) low temperature STM system equipped with an ozone-assisted molecular beam epitaxy (MBE) chamber and the capability of IBA. The flux beam of ozone from a commercial ozone gas delivery system (Fermi Instruments) could be injected into MBE using a nozzle, $\sim$ 40 mm distant from the sample. Optimally doped Bi-2212 single crystals ($T_\textrm{c}$ = 91 K) were grown by a traveling floating zone method \cite{wen2008large}, and \textit{in-situ} cleaved in UHV at room temperature. A following UHV annealing at 250$^\textrm{o}$C leads to a pristine and clean BiO surface of superconducting Bi-2212 [Fig.\ S1], from which the argon-ion bombardment was conducted at an energy of 500 eV and with argon pressure of 1 $\times$ 10$^{-5}$ Torr \cite{supplementary}. Unless otherwise specified, the measurements were performed at 4.2 K using polycrystalline PtIr tips, which were firstly cleaned by \textit{e}-beam heating in MBE and calibrated on Ag/Si(111) films. The STM topography was acquired in a constant-current mode with the bias \textit{V} applied to the sample. Tunneling spectra were measured using a standard lock-in technique with a small bias modulation of 2 meV at 931 Hz.

Figure 1(a) shows an STM topographic image of a Bi-2212 sample prepared by IBA, where atomically flat terraces with various apparent heights are seen \cite{supplementary}. The topographic height distribution, based on a statistical analysis of 30 images, reveals seven apparent peaks, each of which corresponds to an atomic plane [Figs.\ 1(b) and S2]. By comparison of peak-peak separations and the spacings between various crystallographic planes along the \textit{c}-axis [Fig.\ 1(c)], we assign the terraces at different heights as four BiO, two CuO$_2$ and one SrO layer, respectively, as labeled in Fig.\ 1(b). The separation of two neighboring BiO(I) planes are measured to be 15.3 {\AA}, matching excellently with the half-unit-cell thickness (15.35 {\AA}) of Bi-2212. Moreover, the two dominant planes, BiO(I) and SrO(I), are separated by $\sim$ 3.1 {\AA}, close to the theoretical value (2.7 {\AA}). The small discrepancy is primarily of electronic origin, since its value alters with \textit{V}, especially in the occupied states (\textit{V} $<$ 0). It is worth noting that the errors appear comparably small ($<$ 0.5 {\AA}) with reference to the spacing between adjacent Bi-2212 planes ($\geq$ 1.7 {\AA}), thus a misassignment of each plane is unlikely.

\begin{figure}[t]
\includegraphics[width=\columnwidth]{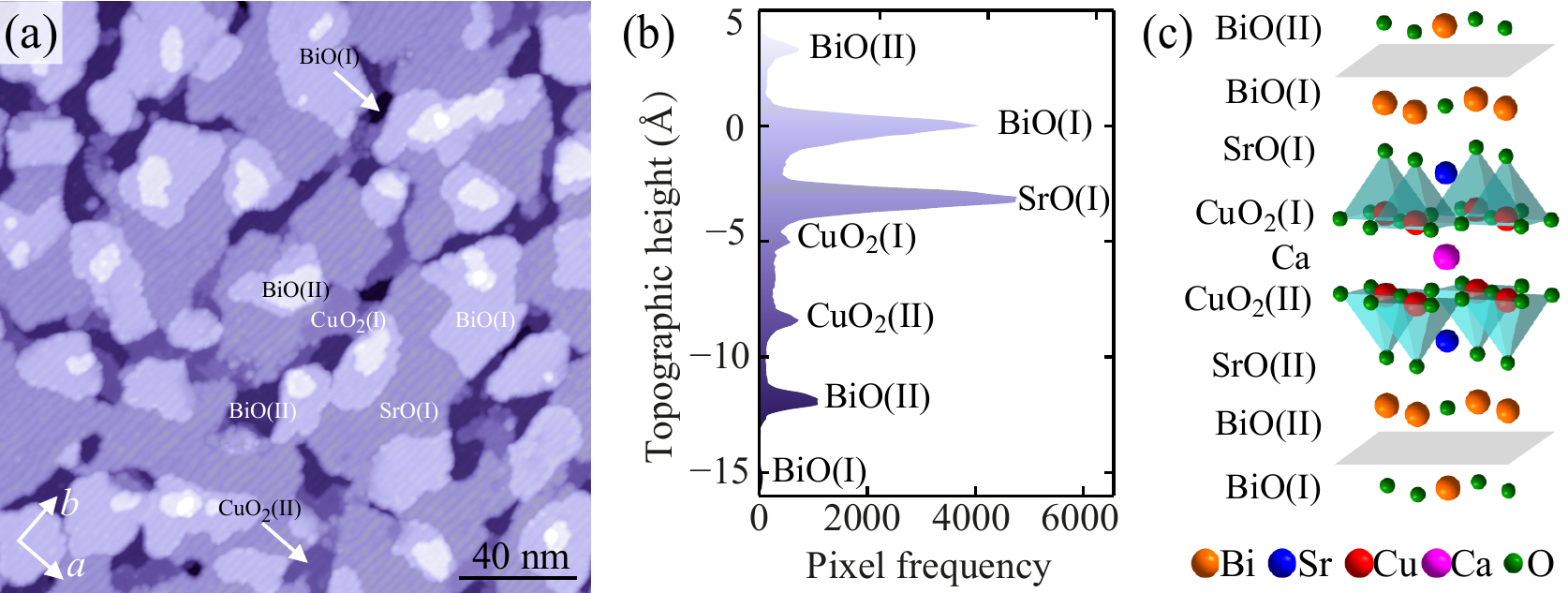}
\caption{(color online) (a) STM topographic image of the as-sputtered and annealed Bi-2212 cuprate, showing various terminating planes ($V$ = 1.5 V, $I$ = 30 pA). The in-plane crystallographic axes ($Cmmm$ space group) are labeled as $a$ and $b$, with $a=b=$ 5.4 {\AA}. (b) Frequency distribution of apparent topographic height, color coded to match those in (a). The zero of topographic height has been deliberately chosen for the top BiO(I) plane. (c) Crystallographic structure of Bi-2212, with repeated inverted oxide layers along the \textit{c}-axis direction. The two gray parallelograms indicate the easily cleaved planes of Bi-2212 crystals.
}
\end{figure}

\begin{figure*}[t]
\includegraphics[width=2\columnwidth]{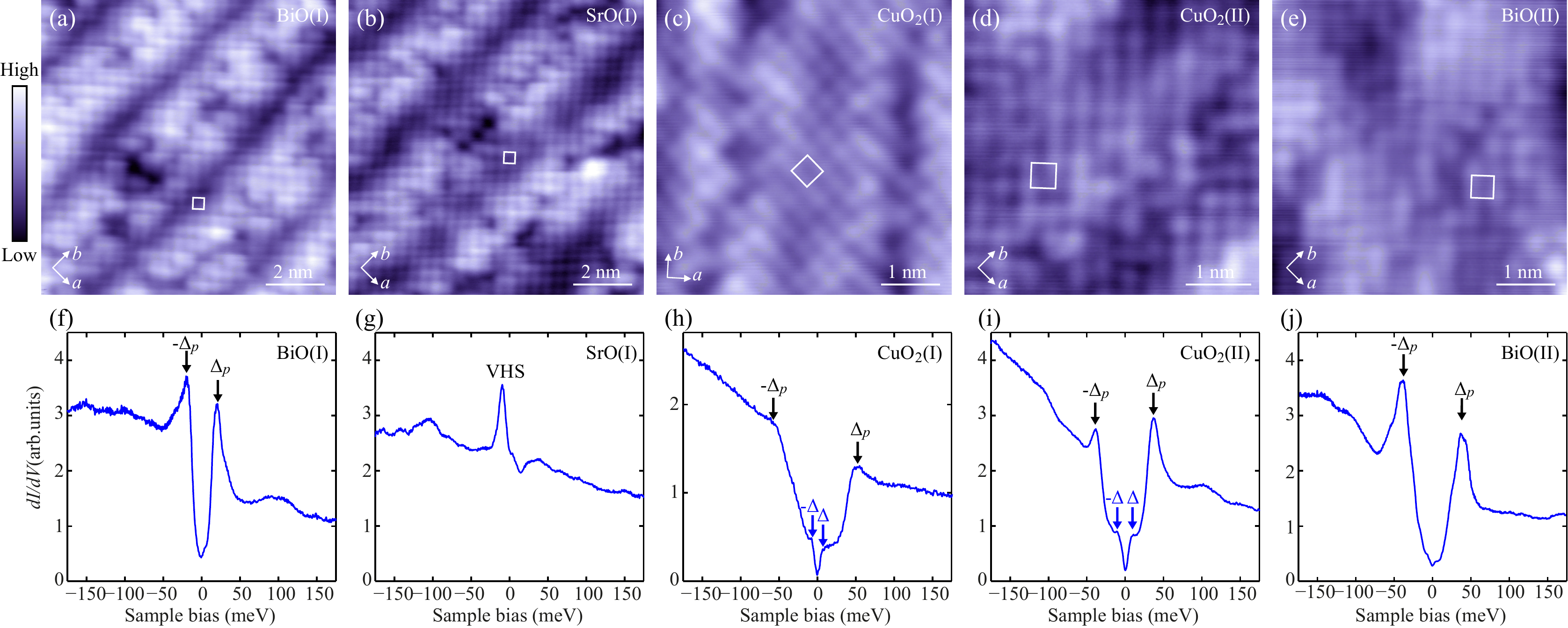}
\caption{(color online) (a-e) Topographies ((a-d) $V$ = 0.2 V, $I$ = 150 pA; (e) $V$ = 0.1 V, $I$ = 200 pA) and (f-j) electronic spectra on various planes of  BiO(I), SrO, CuO$_2$(I), CuO$_2$(II) and BiO(II) in Bi-2212, respectively. Setpoint: $V$ = 0.2 V, $I$ = 400 pA, except for (h) $I$ = 200 pA. White squares mark the respective in-plane unit cells of every exposed planes, with a periodicity of 3.8 {\AA}. Black and blue arrows indicate the PG and superconducting gap in different energy scales, respectively. The VHS on the SrO plane lies near $E_F$.
}
\end{figure*}

Having established the clear-cut identities of all exposed planes of Bi-2212, a question naturally arises as to how the electronic structure changes in different layers and is linked to superconductivity. Prior to answering this question, we first discuss the spectra obtained from the as-sputtered and UHV annealed Bi-2212 samples. As depicted in Fig.\ S3, a broad U-shaped depletion in DOS near $E_F$, but not always symmetric with $E_F$, is observed on the CuO$_2$ planes. This contrasts with the V-shaped DOS suppression at $E_F$ on the BiO planes of the same sample, well-known as ``PG''. The feature, possibly due to a substantial loss of near-surface oxygen dopants during IBA, bears great similarities with those previously reported on the accidentally obtained CuO$_2$(II) planes \cite{misra2002atomic}. The gap asymmetry on the CuO$_2$ planes, together with the anomalous large gap magnitude $>$ 60 meV that exceeds substantially the expected value for the superconductor with $T_\textrm{c}$ = 91 K, points to their nature of neither superconducting gap nor PG. Such results discredit the empirical interpretation of the electronic spectra of BiO layer as the DOS of the buried CuO$_2$ layers which have been reported in previous STM studies, and additionally provide the evidence that the PG might be not intrinsic to CuO$_2$, because it can be easily disturbed in the underdoped regime [Fig.\ S3(b)]. To retrieve and explore the superconductivity-related properties, we annealed the samples under the flux of ozone at 450$^\textrm{o}\textrm{C}$ until the samples re-enter a nearly optimally doped regime with a maximal value of $T_\textrm{c}$ \cite{supplementary}, by comparing with the ``standard'' spectrum of the freshly cleaved superconducting Bi-2212 samples [Fig.\ S1].

Examination of STM topographic images of the recovered superconducting samples reveals the common incommensurate structural buckling feature with a period of $\sim$ 26 {\AA} on all the terminating planes [Figs.\ 2(a-e)], namely the well-known \textit{b}-axis supermodulation in cuprates \cite{fischer2007scanning}. The square lattice with a lattice constant of 3.8 {\AA} on each of the exposed planes could be identified as from the corresponding metal atoms (Bi, Sr or Cu). No surface reconstruction is observed, guaranteeing that the electronic spectra on the exposed BiO, SrO and CuO$_2$ planes are intimately tied to their bulk counterparts in Bi-2212. Shown in the lower panels [Figs.\ 2(f-j)] are the typical differential conductance \textit{dI/dV} spectra, acquired on the different layers shown in the corresponding upper panels, the key features of which exhibit little dependence on the exposed lateral area [Fig.\ S4]. As expected, the BiO(I) layer exhibits the well-defined $E_F$-symmetric PG, while the \textit{dI/dV} spectra of SrO and CuO$_2$ layers come as a surprise [Figs.\ 2(g-i)]. Both CuO$_2$(I) and CuO$_2$(II) layers present the \textit{dI/dV} spectra with two-energy-scale DOS suppression around $E_F$, in contrast to a single PG feature on BiO(I). The PG magnitude $\Delta_p$ of BiO(I) is never reconcilable with either energy scale of the electronic excitations on the CuO$_2$ layers. The results further echo the above claim that the conventional ideology of electronic spectra of BiO(I) plane is questionable.

Further insights into this argument are obtained by spectral measurements of SrO(I) layer that is sandwiched between the BiO(I) and CuO$_2$(I) layers. A pronounced enhancement in DOS near $E_F$, which we interpret as a signature of Van Hove singularity (VHS), is universally found on SrO(I), albeit with some minor site-dependent fine structure [Fig.\ S5]. This discloses a metallic nature of the SrO layer, at odds with the common belief that the SrO is insulating \cite{nieminen2009spectral}. Similar VHS has been previously demonstrated in some regions of Bi$_2$Sr$_2$CuO$_{6+\delta}$ (Bi-2201) cuprate \cite{piriou2011first}, but never in Bi-2212. It has been tentatively accounted for, within a spin fluctuation pairing scenario, as a stronger coupling of VHS to the spin fluctuation in Bi-2212 \cite{piriou2011first, hoogenboom2003modeling}. Our finding suggests that a VHS might be generic but exists only in the SrO layer, and thus has nothing or little to do with the pairing property such as order-parameter symmetry in the superconducting CuO$_2$ layers. In Bi-2212, the prominent PG in the BiO(I) layer hampers a direct visualization of the VHS in the underlying SrO(I) layer. Since the VHS develops only after the Bi-2212 samples were annealed to recover superconductivity under the ozone flux, we argue that it might most probably originate from the interstitial oxygen dopants in the SrO(I) layer \cite{zhou2007correlating, zeljkovic2012imaging}, which shift the valance band of insulating SrO upwards to the $E_F$ \cite{mcleod2010band}. Nevertheless, the metallic nature of SrO(I) layer leaves little possibility that the PG observed on BiO(I) has a simple root at the subsurface CuO$_2$ layer due to a missing of VHS around $E_F$ on BiO(I). Instead we propose that the PG observed by STM might be intrinsic to the BiO layer, which constitutes one of the major findings in our study and will further be discussed below.

\begin{figure}[h]
\includegraphics[width=1\columnwidth]{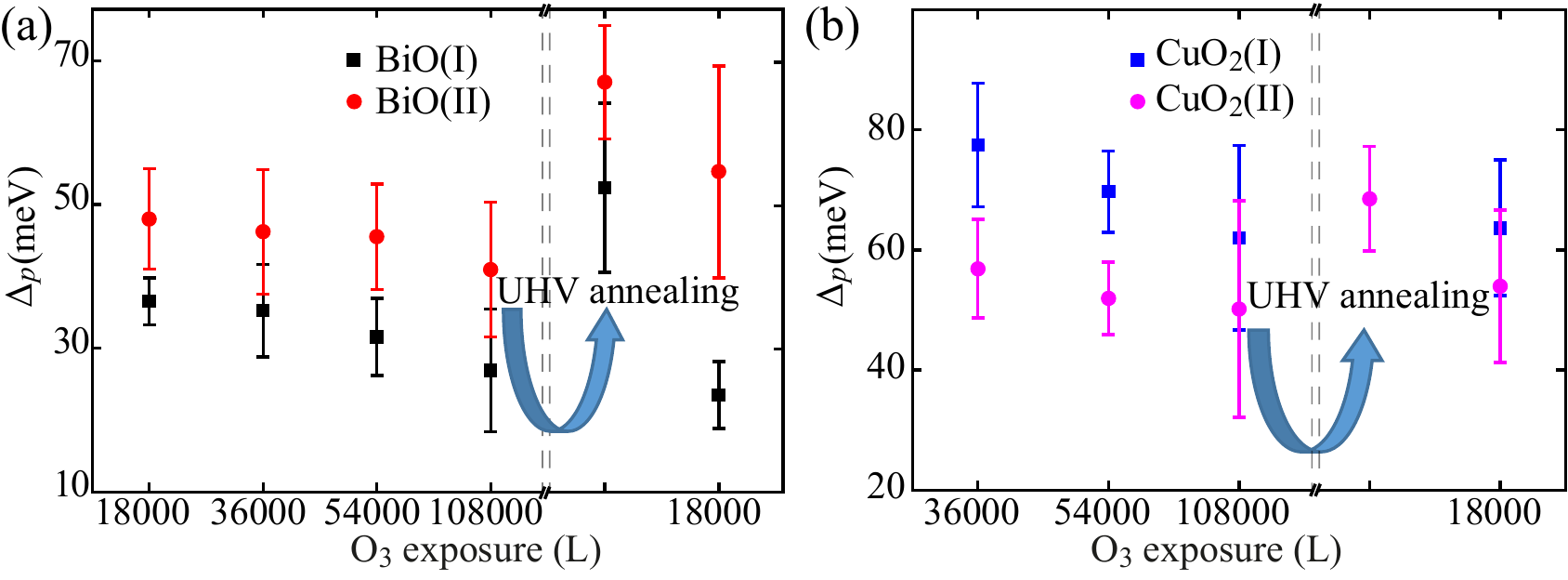}
\caption{(color online) (a, b) Ozone exposure and UHV annealing dependence of $\Delta_p$ on (a) BiO and (b) CuO$_2$ planes. The sequence-annealing under the ozone flux (from left to right) reduces gradually $\Delta_p$, while the annealing in UHV does the opposite. Error bars are estimated as the standard deviations of $\Delta_p$ measured in various conductance spectra. We use langmuri (1 L = 10$^{-6}$ Torr$\cdot$ second) to denote the ozone exposure.
}
\end{figure}

In order to put forward our conjecture more concretely, we explore the electronic spectra of BiO(II) layer [Fig.\ 2(e)], which sits on the top of BiO(I) and binds with BiO(I) via a rather weak van der Waals interaction \cite{matsui1989profile}. This allows for a rational explanation of the spectral feature measured on BiO(II) as more intrinsic to BiO itself. Remarkably the electronic spectrum of BiO(II) layer reveals a clear PG feature again [Fig.\ 2(j)], resembling with those of BiO(I) in a prominent manner. Our statistic of the PG magnitude $\Delta_p$ shows that $\Delta_p$ on BiO(II) is on average larger than that on BiO(I), irrespective of the ozone exposure and UHV annealing [Fig.\ 3(a)]. This contradicts with the empirical model of PG, in which $\Delta_p$ in the adjacent BiO(I) and BiO(II) layers are not expected to change much. Based on our interpretation, the difference in $\Delta_p$ could be understood: besides that both BiO(I) and BiO(II) layers accept interstitial oxygen dopants and become hole-doped \cite{mcelroy2005atomic}, the adjacent SrO might dope more carriers into the BiO(I) layer so that its $\Delta_p$ appears smaller. The interpretation of PG as an intrinsic property of BiO is further corroborated by our experimental observation of an analogous ``PG'' on thicker bismuth oxides grown by oxide-MBE \cite{supplementary}. As seen in Fig.\ 4, the so-called PG can persist on a 4 nm thick Bi oxide island ($\sim$ 13 BiO layers). One may assign the thick bismuth oxide as Bi$_2$O$_3$ or other form of oxides than BiO or their mixture. However, it does not affect our conclusion, rather, gives strong support that the PG may be generic to bismuth oxides. The above results on various BiO layers unambiguously reveal that the PG in Bi-2212 is inherent to the BiO and its relevancy to the superconductivity in Bi-2212 is extremely low.

\begin{figure}[t]
\includegraphics[width=\columnwidth]{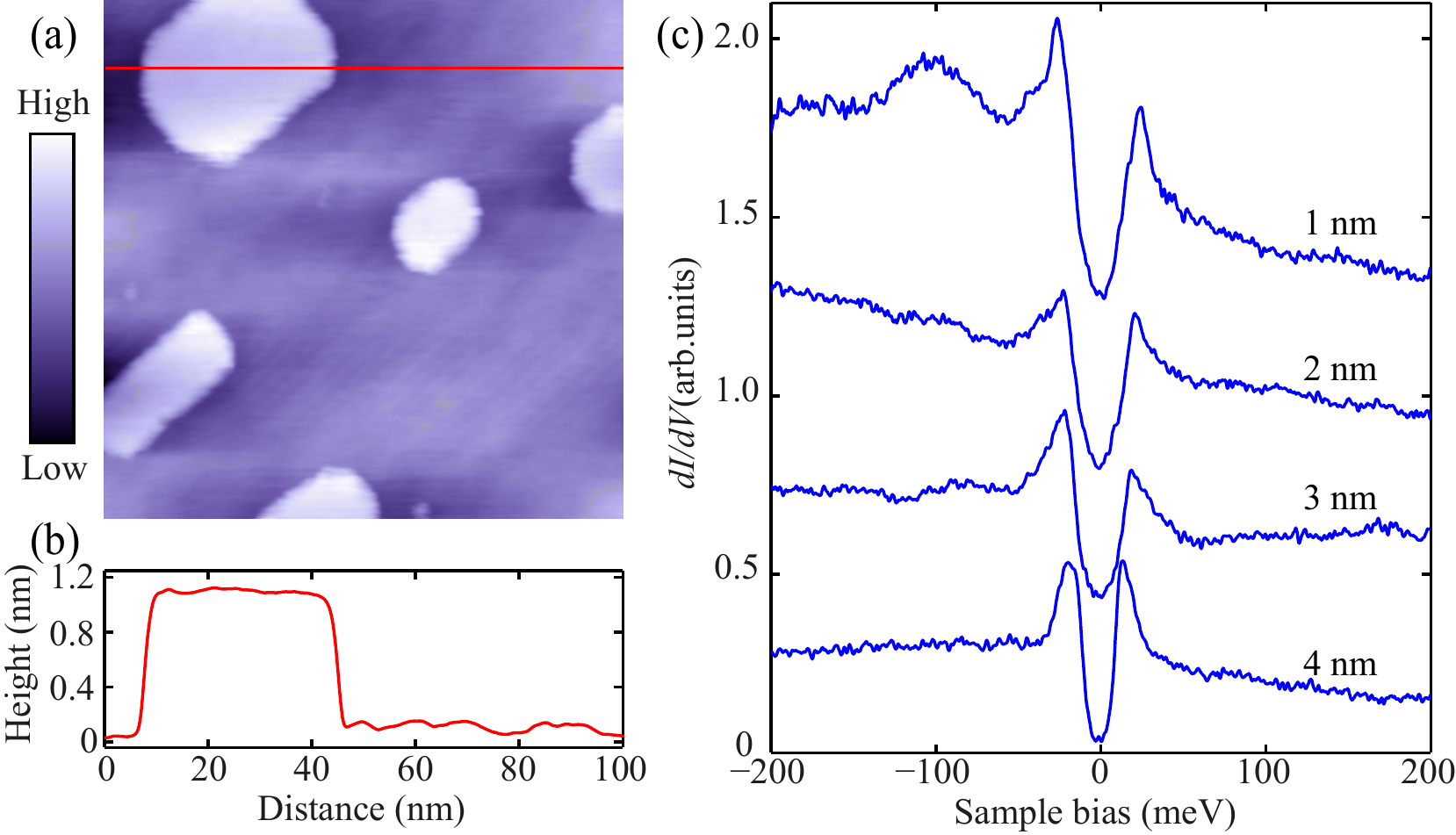}
\caption{(color online) (a) STM topography showing atomically flat BiO$_x$ islands on Bi-2212 grown by MBE ($V$ = 6.0 V, $I$ = 30 pA, 100 nm $\times$ 100 nm). (b) Height profile taken along the red line cut indicated in (a), revealing the BiO$_x$ island with a thickness of $\sim$ 1 nm.  (c) Typical \textit{dI/dV} spectra on the BiO$_x$ islands with various thicknesses (Setpoint: $V$ = 0.2 V, $I$ = 50 pA), indicating a PG with $\Delta_p\sim26\pm7$ meV.
}
\end{figure}

For completeness, we revisit the electronic spectra on CuO$_2$, the major building layers of cuprates. Figure 5(a) displays a series of \textit{dI/dV} spectra along a 4-nm trajectory on a CuO$_2$(II) plane, confirming the two characteristic energy scales in DOS near $E_F$. Such a coexisting two-gap feature was sometimes - but not always - seen in a few superconducting La$_{\textrm{2-}x}$Sr$_x$CuO$_4$ and Bi-based cuprates \cite{kato2008doping, tacon2006two, boyer2007imaging}, giving rise to a fierce debate over their respective nature \cite{hufner2008two}.  The larger energy gaps, which appear larger in CuO$_2$(I) than in CuO$_2$(II) layer due to the pivotal role of the adjacent SrO layer, shares a similar behavior to the PG of BiO [Fig.\ 3]: the higher the hole concentration is, the smaller the gap magnitude $\Delta_p$ is. This suggests that they may either link to the spectra of BiO (opposite to the common wisdom) or originate from their respective property. Although our observation that the PGs on BiO are more well-defined (substantial accumulation of spectral weight at the gap edges) than those on CuO$_2$ [Fig.\ 2] seems to support the former scenario, we cannot exclude the latter: the PG may be generic to metal oxides \cite{mannella2005nodal, kim2014fermi, kim2015observation}. Nevertheless, our results indicate that there should be very little relevance between the PG and superconductivity in cuprates. Based on this standpoint, one can understand the longstanding puzzle of the completely different behaviors of the PG and $T_\textrm{c}$ versus hole concentration in the phase diagram of cuprates: the linearly dependent PG is mainly a measure of the conductivity (reflects the availability of holes) of the BiO layer while the $T_\textrm{c}$ or superconducting gap with dome-like dependence reflects the amount of holes actually participating in Cooper pairing in the CuO$_2$ layer \cite{hufner2008two}.

\begin{figure}[t]
\includegraphics[width=1\columnwidth]{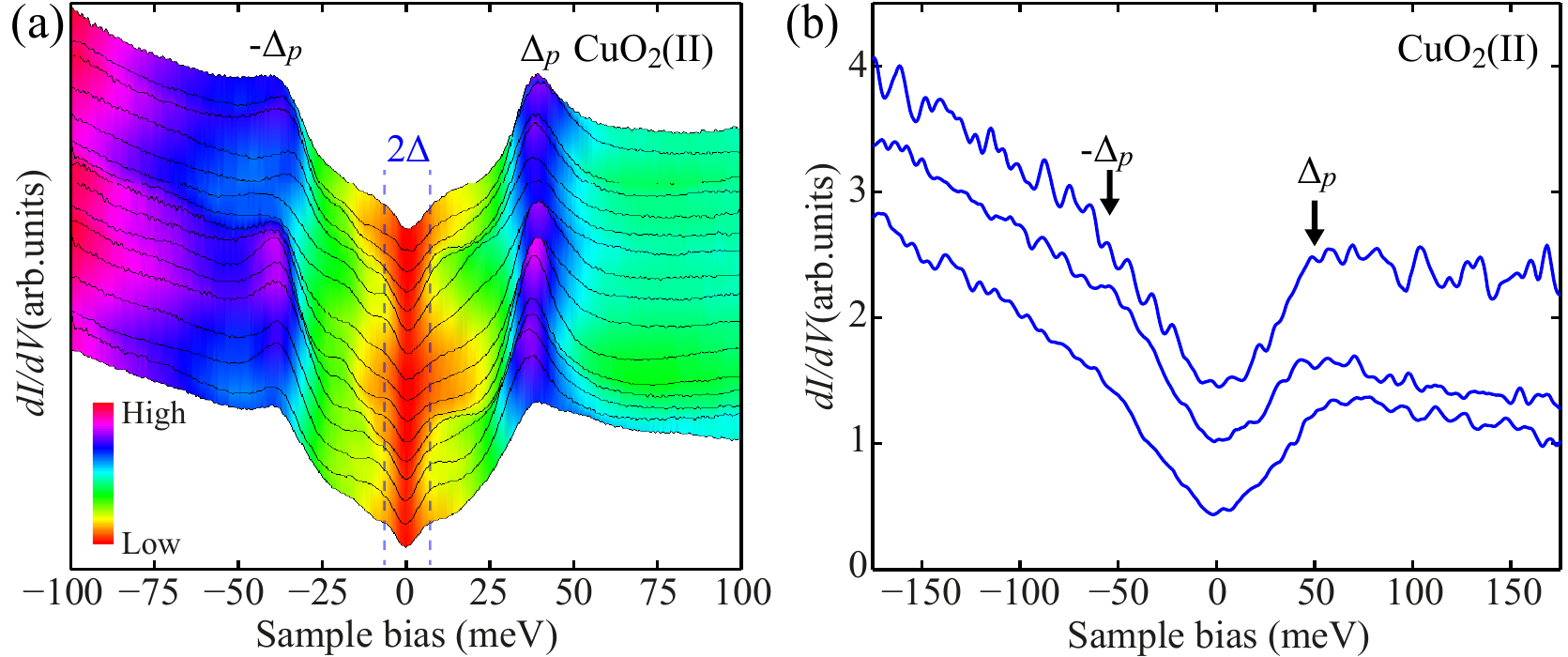}
\caption{(color online) (a) Differential conductance $dI/dV$ spectra acquired along a 4-nm trajectory on a CuO$_2$(II) plane at 4.2 K. The two vertical dashes indicate both edges of the superconducting gap. Setpoint: $V$ = 0.2 V, $I$ = 400 pA. (b) Three $dI/dV$ spectra on distinct CuO$_2$(II) planes at 78 K. Setpoint: $V$ = 0.2 V, $I$ = 100 pA. The spectra have been vertically offset for clarity.
}
\end{figure}

More significantly, our finding - the robust smaller energy scale gap ($\Delta$) on the superconducting CuO$_2$ planes [Fig.\ 5(b)] - hints that $\Delta$ is most likely the real and the only superconducting gap. This is strongly evidenced by the invisibility of the smaller-energy-scale gap at a higher measurement temperature of 78 K (close to $T_\textrm{c}$), whereas the larger pseudogap remains prominent. By measuring $\Delta$ $\sim 15\pm4$ meV and using $T_\textrm{c}$ = 91 K of optimally-doped Bi-2212 sample, we estimate the reduced gap magnitude $2\Delta/k_\textrm{B}T_\textrm{c}$ = 3.8 $\pm$ 1.0, which is in line with the value (3.53) in BCS theory.

Our systematic atomic-layer-resolved STM study has demonstrated distinctive differences in the electronic structures of ingredient oxide layers in Bi-2212. This study, for the first time, enables us to address the respective role played by each oxide layer in Bi-2212 in an unprecedented way. The most studied BiO layers, as well as the thicker bismuth oxide films, exhibit the well-known PG feature, which we demonstrate is irrelevant to the superconductivity property of Bi-2212. The BiO layers, together with CuO$_2$ and SrO layers, may mainly serve as framework to establish the perovskite crystal structure and chemical stoichiometry of cuprates. The SrO layer, with enhanced DOS near $E_F$ due to the VHS, is revealed crucial and acts carrier reservoir for the adjacent CuO$_2$ layers to boost the superconductivity therein. Under this context, the superconductivity of Bi-2212, by implication all other cuprate superconductors, may not be as complicated as anticipated: like FeSe/SrTiO$_3$, the CuO$_2$/SrO bilayer might hold the key ingredients for realizing high-$T_\textrm{c}$. Our study suggests that in order to eventually understand the paring mechanism of cuprates and clarify the existing controversies, preparation of CuO$_2$ superconducting layers and direct measurement of their electronic structure are essential, which are currently under way.

\begin{acknowledgments}
This work was financially supported by National Science Foundation and Ministry of Science and Technology of China. Work at Brookhaven was supported by the Office of Basic Energy Sciences (BES), Division of Materials Sciences and Engineering, U. S. Department of Energy (DOE), through Contract No. DE-SC00112704.
\end{acknowledgments}

%\bibliography{BSCCO}
%

%%%%%%%%%% Merge with supplemental materials %%%%%%%%%%
% \pagebreak
\widetext
\newpage
%%%%%%%%%% Merge with supplemental materials %%%%%%%%%%
%%%%%%%%%% Prefix a "S" to all equations, figures, tables and reset the counter %%%%%%%%%%
\setcounter{equation}{0}
\setcounter{figure}{0}
\setcounter{table}{0}
\setcounter{page}{5}
\makeatletter
\renewcommand{\theequation}{S\arabic{equation}}
\renewcommand{\thefigure}{S\arabic{figure}}
\renewcommand{\bibnumfmt}[1]{[S#1]}
\renewcommand{\citenumfont}[1]{S#1}
%%%%%%%%%% Prefix a "S" to all equations, figures, tables and reset the counter %%%%%%%%%%

\onecolumngrid
\Large
{\textbf{Supplemental Material for:} }
\begin{center}
\textbf{\large Mapping the Electronic Structure of Each Ingredient Oxide Layer of High-$T_\textrm{c}$ Cuprate Superconductor Bi$_2$Sr$_2$CaCu$_2$O$_{8+\delta}$}
\end{center}
\small
\maketitle
\twocolumngrid
\begin{spacing}{1.6}
Our oxide MBE-STM combined system has a base pressure of better than 1 $\times$ 10$^{-10}$ Torr. The energy resolution of $dI/dV$ spectrum is better than 0.1 meV at 400 mK \cite{ji2008high}. The pixels for all STM topographic images are 512 $\times$ 512. All STM images were processed using WSxM software \cite{horcas2007wsxm}.

\begin{figure}[h]
\includegraphics[width=0.76\columnwidth]{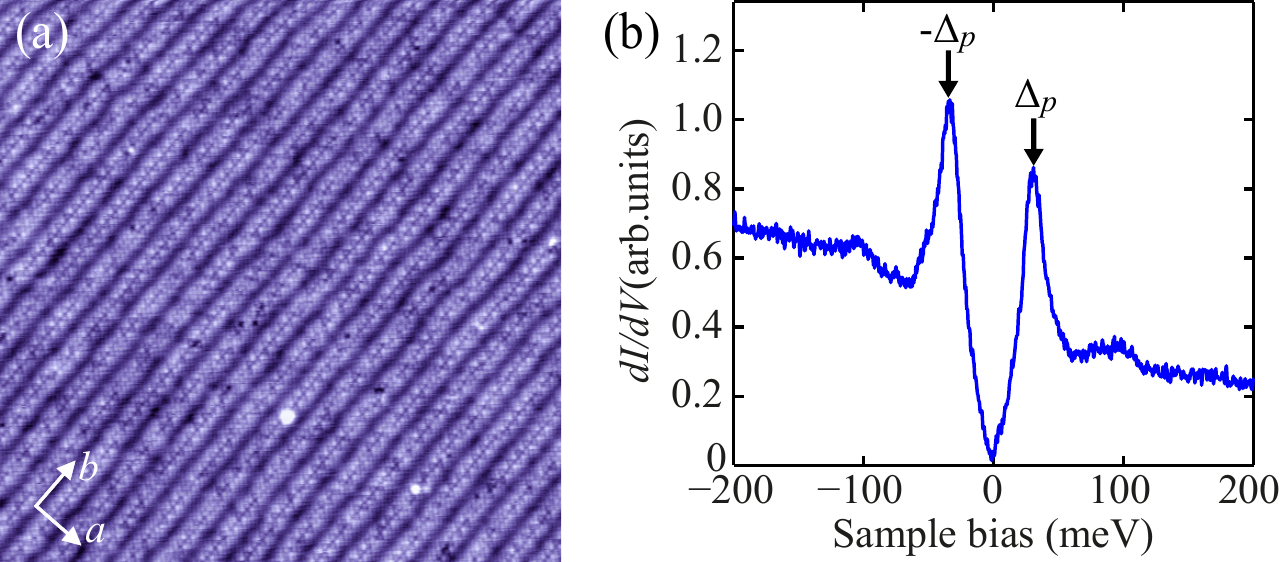}
\caption{(a) Atomically resolved STM topographic image ($V$ = 1.5 V, $I$ = 25 pA, 50 nm $\times$ 50 nm) and (b) \textit{dI/dV} spectrum taken on the pristine BiO surface of vacuum cleaved superconducting Bi-2212 crystal ($T_\textrm{c}$ = 91 K). The PG $\Delta_p\sim$ 40 meV is visible. The tunneling conditions: $V$ = 0.2 V, $I$ = 100 pA.
}
\end{figure}
\begin{figure}[h]
\includegraphics[width=\columnwidth]{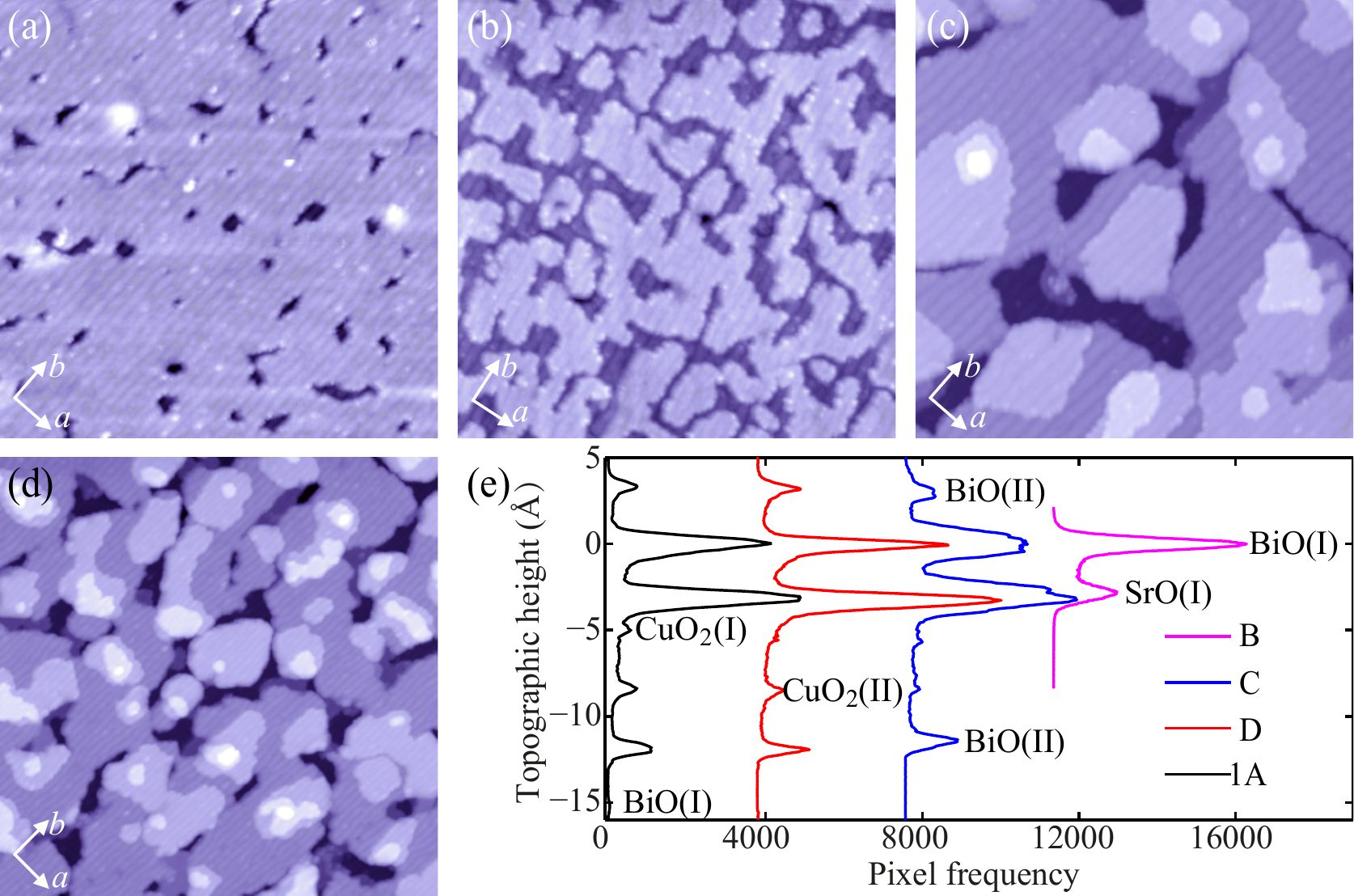}
\caption{(a-d) The topography evolution of Bi-2212 with increasing cycles of IBA ($V$ = 1.5 V, $I$ = 30 pA). The size of the images is 100 nm $\times$ 100 nm in (a-c) and 200 nm $\times$ 200 nm in (d). (e) Frequency distribution of topographic height in (b-d) as well as in Fig.\ 1(a), ascertaining the identities of the terminating planes. The curves have been horizontally offset for clarity.
}
\end{figure}

\begin{figure}[h]
\includegraphics[width=0.9\columnwidth]{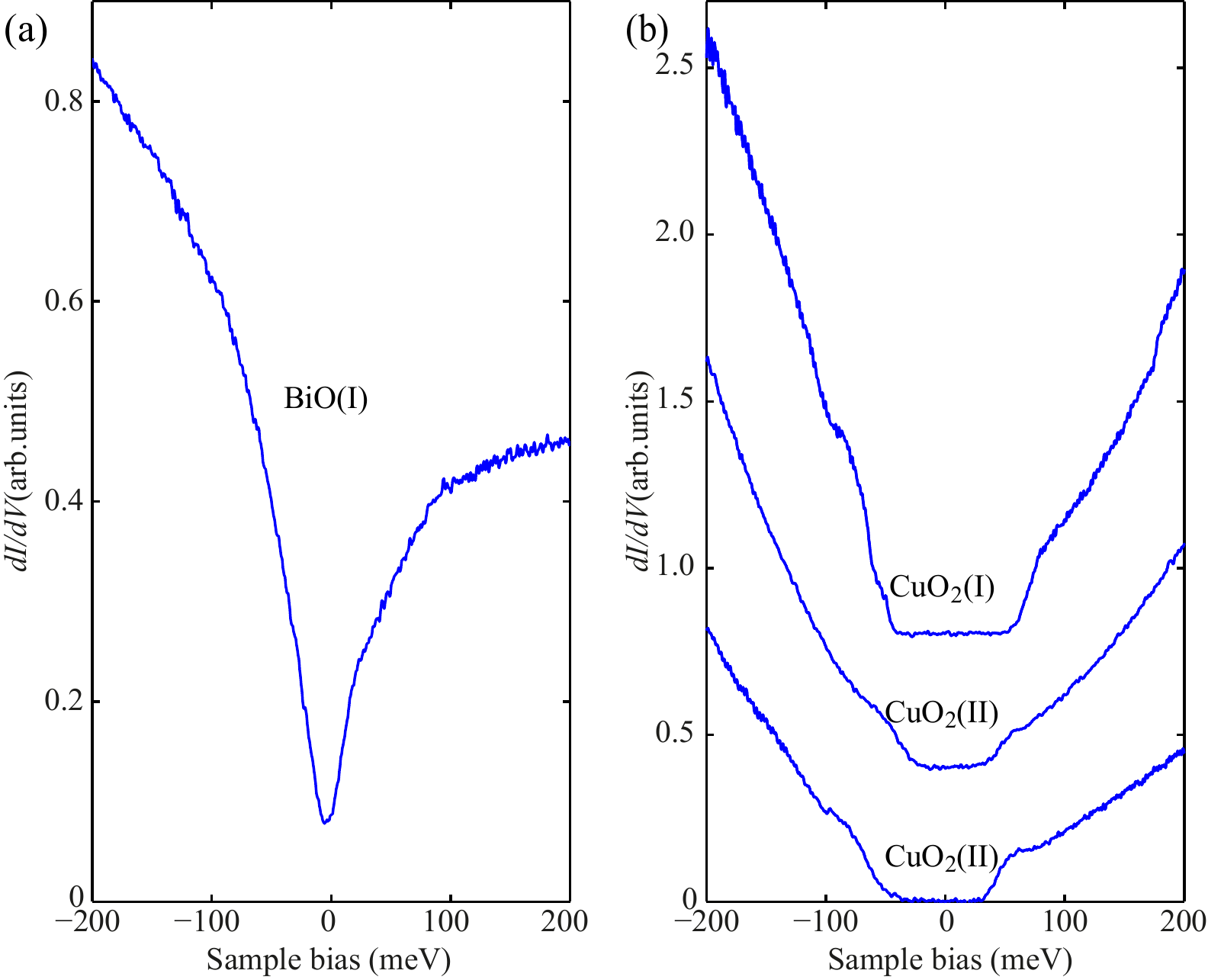}
\caption{(a, b) STM spectra measured on (a) BiO(I), (b) CuO$_2$(I) and CuO$_2$(II) planes. The sample is the as-sputtered and UHV annealed Bi-2212 single crystal. The curves in (b) have been vertically offset for clarity. The fact that the PG occurred to be anomalously large ($\Delta_p>$ 100 meV) on BiO(I), together with the vanishing pronounced peaks at the gap edges, indicates the sample is in underdoped regime. Setpoint: $V$ = 0.2 V, $I$ = 50 pA.
}
\end{figure}

\begin{figure*}[t]
\includegraphics[width=1.57\columnwidth]{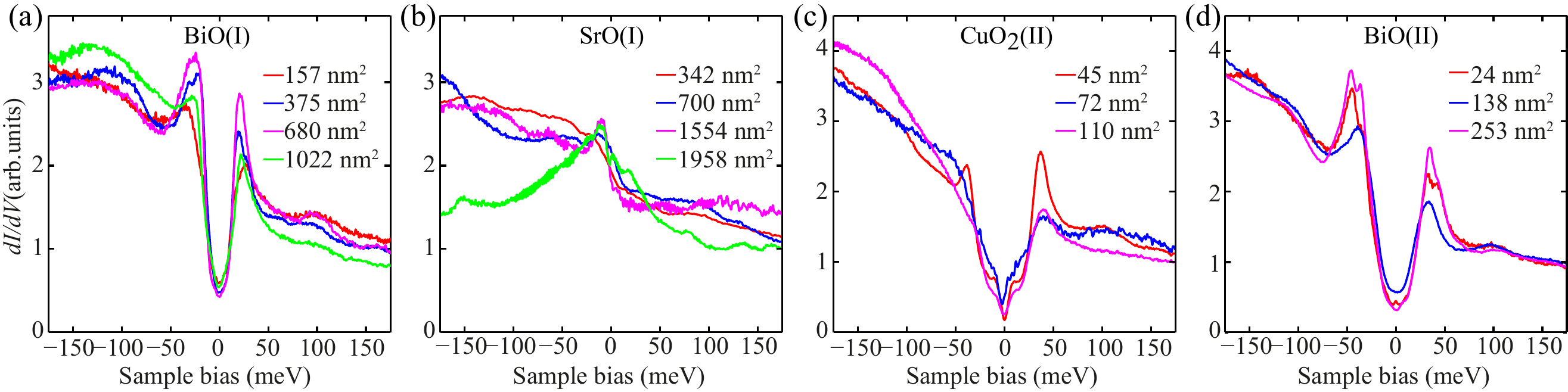}
\caption{Lateral area $A$-independent $dI/dV$ spectral features on the exposed (a) BiO(I), (b) SrO(I), (c) CuO$_2$(II) and (d) BiO(II) planes, respectively. The small discrepancies in the fine structure might originate from an intrinsic inhomogeneity of cuprate superconductors. The tunneling conditions: $V$ = 0.2 V, $I$ = 400 pA.
}
\end{figure*}

\begin{figure}[h]
\includegraphics[width=0.8\columnwidth]{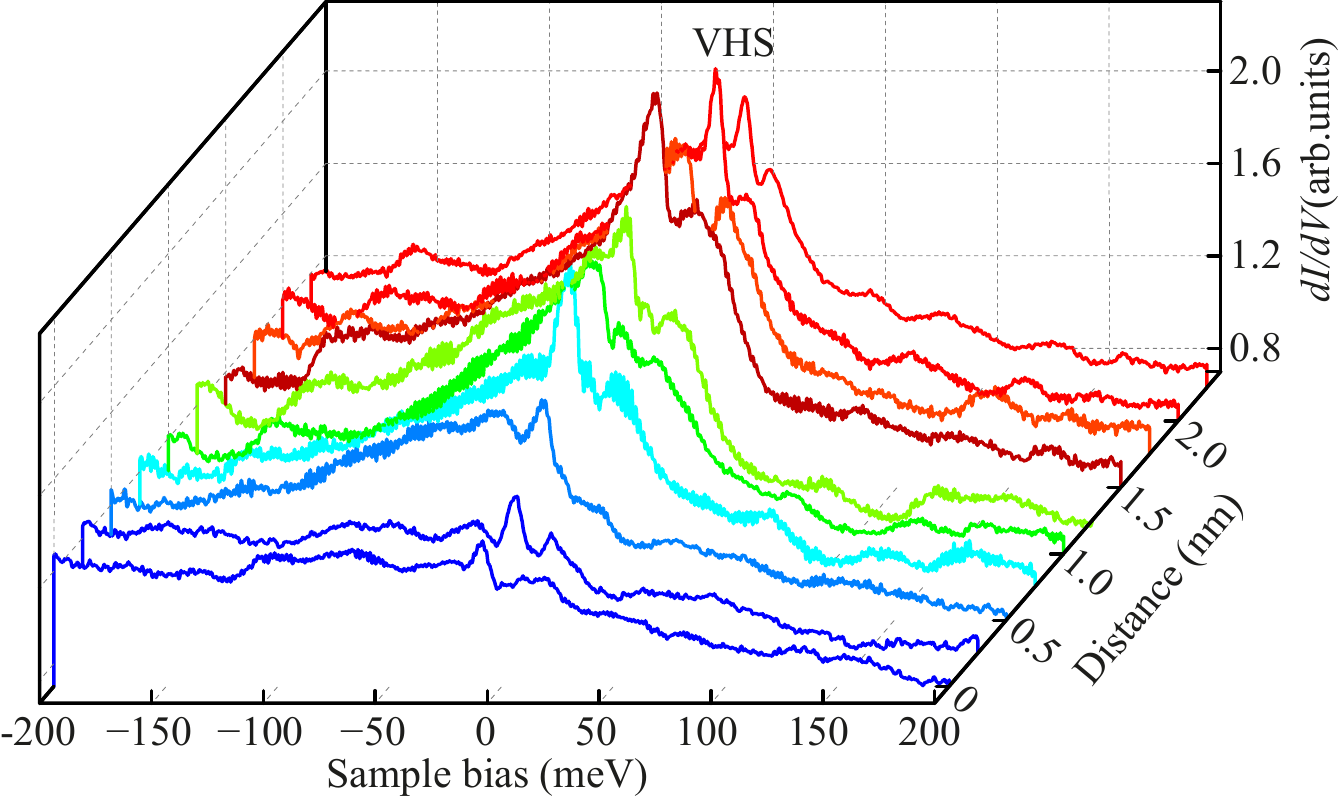}
\caption{(color online) STM spectra acquired along a 2.5-nm trajectory on the SrO(I) plane. Despite subtle changes in the fine structure, every spectrum presents the appreciable DOS enhancement at around $-$21 $\pm$ 13 meV, indicative of a general VHS near $E_F$ on the SrO plane. Setpoint: $V$ = 0.2 V, $I$ = 400 pA.
}
\end{figure}

The as-sputtered Bi-2212 single crystal exhibits a rather disordered surface. Our subsequent annealing under UHV condition at 500$^\textrm{o}$C leads to atomically flat surface. Figure S2 illustrates the IBA effects on the geometry and surface termination of sputtered Bi-2212. The annealing under the ozone flux beam (from 1 $\times$ 10$^{-5}$ Torr to 3 $\times$ 10$^{-5}$ Torr) was conducted at an optimal sample temperature of 450$^\textrm{o}$C. A higher or lower temperature is found to lead to a relatively lower efficiency of oxygen dopant incorporation and thus poor superconducting property of the Bi-2212 sample. With increasing oxygen interstitial dopants by this optimized annealing procedure under the ozone flux, the Bi-2212 single crystal can evolve progressively into the underdoped, nearly optimally doped and even over-doped regions [Fig.\ 3].

Bismuth oxides BiO$_x$ with unidentified stoichiometry of bismuth and oxygen were prepared by depositing Bi under a flowing ozone flux of 1 $\times$ 10$^{-5}$ Torr at 300$^\textrm{o}$C, followed by further exposure of ozone without Bi flux at 450$^\textrm{o}$C. Depending on the Bi coverage, the islands of bismuth oxides range from 1 nm to 4 nm in height. The differential conductance spectra on the islands invariably reveal a PG-like DOS depression around $E_F$. Moreover, the PG is spatially rather homogeneous, similar to those observed on the BiO surface of the overdoped cuprate superconductors (Fig.\ 4).
\end{spacing}

%\bibliography{BSCCO}

\begin{thebibliography}{30}
\bibitem [{sup()}]{supplementary}%
  \BibitemOpen
  \href@noop {} {\bibinfo  {journal} {Materials and Method are available as
  supporting material}\ }\BibitemShut {NoStop}%
\end{thebibliography}
%
\end{document}